\documentclass[twoside,fleqn]{article}

\def\noi{\noindent}

\makeatletter

\renewcommand{\thesubsubsection}%
        {\arabic{section}.\arabic{subsection}.\arabic{subsubsection}.}
\renewcommand{\@oddhead}{\raisebox{0pt}[\headheight][0pt]{%
   \vbox{\hbox to\textwidth{\rightmark \hfil \rm \thepage \strut}\hrule}}}
\renewcommand{\@evenhead}{\raisebox{0pt}[\headheight][0pt]{%
   \vbox{\hbox to\textwidth{\thepage \hfil \leftmark \strut}\hrule}}}
\newcommand{\heads}[2]{\markboth{\protect\small\it #1}{\protect\small\it #2}}
\newcommand{\Acknow}[1]{\subsection*{Acknowledgement} #1}
\makeatother

\newcommand{\Title}[1]{\noi {\Large #1} \\}
\newcommand{\Author}[2]{\noi{\large\bf #1}\\[2ex]\noi{\it #2}\\}
\newcommand{\Abstract}[1]{\vskip 2mm \begin{center}
        \parbox{16.4cm}{\small\noi #1} \end{center}\bigskip}
\newcommand{\foom}[1]{\protect\footnotemark[#1]}
\newcommand{\foox}[2]{\footnotetext[#1]{#2}\addtocounter{footnote}{1}}

\def\sect{Sec.\,}
\def\ssect#1{Subsec.\,#1}

\def\nq{\hspace{-1em}}
\def\nqq{\hspace{-2em}}
\def\nhq{\hspace{-0.5em}}

\def\cm{\hspace{1cm}}
\def\inch{\hspace{1in}}
\def\mas{\mbox{$\mathstrut$}}

\def\nhx{\vspace*{-0.5ex}}

\def\eq{Eq.\,}
\def\eqs{Eqs.\,}
\def\beq{\begin{equation}}
\def\eeq{\end{equation}}
\def\bear{\begin{eqnarray}}
\def\al{&\nhq}
\def\lal{&&\nqq {}}               
\def\bearr{\begin{eqnarray} \lal}
\def\ear{\end{eqnarray}}
\def\earn{\nonumber \end{eqnarray}}
\def\tst{\textstyle}
\def\dst{\displaystyle}

\def\nnnv{\nonumber\\[5pt] \lal }
\def\yy{\\[5pt]}

\def\eql{\al =\al}

\def\eqos{\ \stackrel{\rm OS}{=}\ }
\def\eqdef{\stackrel{\rm def}{=}}
\def\e{{\,\rm e}}
\def\d{\partial}

\def\sign{\mathop{\rm sign}\nolimits}

\def\dim{\mathop{\rm dim}\nolimits}
\def\const{{\rm const}}

\def\half{{\tst\frac{1}{2}}}

\newcommand{\vars}[1]{\left\{\begin{array}{ll}#1\end{array}\right.}

\def\ep{\epsilon}
\def\o{\omega}
\def\A{{\cal A}}

\def\M{{\cal M}}

\def\S{{\cal S}}
\def\V{{\cal V}}
\def\TH{T_{\rm H}}
\def\oI{{\overline I}}
\def\od{{\overline d}}
\def\oD{{\overline D}}
\def\olam{\overline{\lambda}}
\def\ophi{\overline{\varphi}}
\def\uc{{\underline c}}
\def\vY{\vec Y{}}
\def\vZ{\vec Z{}}
\def\hq{\hat q{}}
\def\Som{\S_{\o}}
\def\sumo{\sum_{\o}}
\def\summo{\sum_{\mu\in\Som}}
\def\sumn{\sum_{i=1}^{n}}
\def\sums{\sum_s}
\def\umx{u_{\max}}
\def\m{{\rm m}}

\def\Nsq{N_s^2}
\def\Noq{N_\o^2}
\def\Fei{F_{\e I}}
\def\Fmi{F_{\m I}}
\def\Qei{Q_{\e I}}
\def\Qmi{Q_{\m I}}

\newcommand{\eps}{ \varepsilon }
\newcommand{\R}{{\sf R\hspace*{-0.9ex}\rule{0.1ex}{1.5ex}\hspace*{0.9ex}}}

\def\Nsq{N_s^2}

\def\rank{\mathop{\rm rank}\nolimits}
\def\SE{S_{\mbox{\small E}}}

\def\sph{spherically symmetric\ }
\def\bh{black hole}
\def\bhs{black holes}
\def\wh{wormhole}
\def\whs{wormholes}

\heads{K.A. Bronnikov}{Block-Orthogonal Brane Systems, Black Holes and
       Wormholes}

\begin{document}
\thispagestyle{empty}
    \unitlength=1mm \begin{picture}(166,15)
        \put(126,8){\shortstack[r]{{\bf PDF-UFES 003/97}\\
				   {\bf hep-th/9710207}    }}
    \end{picture}

\Title
{\LARGE\bf Block-orthogonal brane systems,\yy black holes and wormholes}

\Author{K.A. Bronnikov}
{Departamento de F\'{\i}sica, Universidade Federal do Esp\'{\i}rito Santo,
Vit\'oria --- CEP 29060--900, ES, Brazil\foom 1}

\Abstract
     {Multidimensional cosmological, static \sph and Euclidean
     configurations are described in a unified way for gravity interacting
     with several dilatonic fields and antisymmetric forms,
     associated with electric and magnetic $p$-branes.
     Exact solutions are obtained when certain
     vectors, built from the input parameters of the model, are either
     orthogonal in the minisuperspace, or form mutually orthogonal
     subsystems. Some properties of \bh\ solutions are indicated, in
     particular, a no-hair-type theorem and restrictions emerging in models
     with multiple times. From the non-existence of Lorentzian \whs, a
     universal restriction is obtained, applicable to orthogonal or
     block-orthogonal subsystems of any $p$-brane systems. Euclidean \wh\
     solutions are found, their actions and radii are explicitly
     calculated.}

\centerline{PACS numbers: 04.40, 04.50, 04.70}

\foox 1 {E-mail: kb@cce.ufes.br\\
Permanent address: Centre for Gravitation and Fundamental Metrology,
VNIIMS, 3-1 M. Ulyanovoy St., Moscow 117313, Russia, e-mail:
kb@rgs.mccme.ru}

\section{Introduction. The model} 

     Multiple self-gravitating dilatonic fields and antisymmetric
     forms, associated with $p$-branes, naturally emerge in
     bosonic sectors of supergravities \cite{SS}, superstring
     and M-theory, their generalizations and modifications [2--5].
     This paper continues the recent studies of such models begun in Refs.
     [6--11].

     For fields depending on a single coordinate,
     we present in \sect 2 a general exact solution, assuming that the
     characteristic vectors $\vY_s$ built from the input parameters,
     form an orthogonal system (OS) in minisuperspace. This solution
     generalizes many previous ones (\cite{LPTX,br95,AIV}, etc.)
     and was first found in its present form in \cite{bim97}.
     A new class of exact solutions is built for systems where
     $\vY_s$ are not all orthogonal, but form a block-orthogonal system
     (BOS). The OS solution is included here as a special case.
     \sect 3 discusses \bh\ solutions; it is shown, in particular, that
     even in spaces with multiple times a black hole may only exist with
     its unique, one-dimensional time. In \sect 4 the absence of Lorentzian
     \whs\ for  fields with positive energy is used to obtain a
     universal restriction on the $p$-brane system parameters. Unlike
     Lorentzian \whs, Euclidean ones do exist and are briefly
     discussed in \sect 5. A few examples are given in \sect 6.

     We consider $D$-dimensional gravity interacting with several
     antisymmetric $n_s$-forms $F_s$ and dilatonic scalar fields
     $\varphi^a$, with the action
\beq                                                       \label{2.1}
     S = \frac{1}{2\kappa^{2}}
	             \int\limits_{\M} d^{D}z \sqrt{|g|} \biggl\{
	R[g]
	- \delta_{ab} g^{MN} \d_{M} \varphi^a \d_{N} \varphi^b
                                                    - \sum_{s\in \S}
 	\frac{\eta_s}{n_s!} \e^{2 \lambda_{sa} \varphi^a} F_s^2
                  \biggr\},
\eeq
     in a (pseudo-)Riemannian manifold
     $\M = \R_1(u) \times \M_{0} \times \ldots \times \M_{n}$
     with the metric
\beq
     ds^2 = g_{MN}ds^M dz^N =
            w \e^{2{\alpha}(u)} du^2 +                    \label{2.11}
    	        \sum_{i=0}^{n} \e^{2\beta^i(u)} ds_i^2, \cm  w=\pm 1,
\eeq
     where $u$ is a selected coordinate ranging in $\R_1(u) \subseteq \R$;
     $g^i = ds_i^2$ are metrics on $d_i$-dimensional
     factor spaces $\M_i$ of arbitrary signatures $\eps_i=\sign g^i$;
     $|g| = |\det g_{MN}|$ and similarly for subspaces;
     $F_s^2 =  F_{s,\ M_1 \ldots M_{n_s}} F_s^{M_1 \ldots M_{n_s}}$;
     $\lambda_{sa}$ are coupling constants;
     $\eta_s = \pm 1$ (to be specified later);
     $s \in \S$,  $a\in \A$, where $\S$ and $\A$ are finite sets.
     All $\M_i$, $i>0$ are assumed to be Ricci-flat,
     while $\M_0$ is a space of constant curvature $K_0= 0,\ \pm 1$.

     We assume $\varphi^a = \varphi^a(u)$ and use the
     indices $s\in \S$ to jointly describe $u$-dependent
     electric ($\Fei$) and magnetic ($\Fmi$) $F$-forms, associated with
     subsets $I = \{ i_1, \ldots, i_k \}$ ($i_1 < \ldots < i_k$) of the set
     of factor space numbers $\{i\} = I_0 = \{0, \ldots, n\}$:
     $\S = \{\e I\} \cup \{\m I\}$.
     Given a set of potential functions $\Phi_s(u)$,
     electric and magnetic $F$-forms are defined for each $I$ as follows:
\beq
     F_{\e I}= d\Phi_{\e I} \wedge \tau_I,   \cm        \label{2.17}
     F_{\m I} = \e^{-2\lambda_{\m Ia}\varphi^a}
	                          * [d\Phi_{\m I}\wedge\tau_I],
\eeq
    where $*$ is the Hodge duality operator,
    $\tau_I \eqdef \tau_{i_1}\wedge \ldots \wedge \tau_{i_k}$,
    $\tau_i$ are the volume forms of $\M_i$,
    $d\Phi = \dot\Phi\, du$ and a dot denotes $d/du$. By construction,
\beq
    n_{\e I}=\rank F_{\e I} = d(I) + 1,\cm\quad
    n_{\m I}=\rank F_{\m I} = D- \rank F_{\e I} = d(\oI),  \label{2.22}
\eeq
    where $	\oI \eqdef I_0 \setminus I$ and
    $d(I) = \sum_{i\in I} d_i$, the dimension of
    $\M_I = \M_{i_1} \times \ldots \times \M_{i_k}$.

    Nonzero components of $\Fei$ carry coordinate indices of the
    subspaces $\M_i,\ i\in I$, those of $\Fmi$ --- indices of
    $\M_i,\ i\in \oI$. In $p$-brane studies \cite{brane}
    it is usually supposed that one of the coordinates of $\M_I$
    is time and a form (\ref{2.17}) corresponds to an electric or magnetic
    $(d(I)-1)$-brane ``living'' in the remaining subspace of $\M_I$.

    Several, instead of one, electric and/or magnetic forms might be
    attached to each $I$; this would change actually nothing in the
    solution process but complicate the notations.

    Our problem setting covers various classes of models:
({\bf A})
    isotropic and anisotropic cosmologies, where $u$
    is timelike, $w=-1$;
({\bf B})
    static models with various symmetries (spherical, planar,
    etc.), where $u$ is spacelike, $w=+1$ and time is selected
    among $\M_i$;
({\bf C})
    Euclidean models with similar symmetries, or models
    with a Euclidean ``external" space-time, $w=+1$.

    In all Lorentzian models with the signature $(-++\ldots +)$, the
    energy density $-T^t_t$ of the fields $F_s$ is non-negative
    if one chooses in (\ref{2.1}), as usual, $\eta_s = 1$ for all $s$.
    In models with arbitrary $\eps_i$, one obtains $-T^t_t \geq 0$ if
\beq
     \eta_{\e I} = - \eps(I)\eps_t(I),\cm
     \eta_{\m I}= - \eps(\oI) \eps_t(\oI),  \cm\cm   \label{2.25}
	\eps(I)\eqdef \prod_{i\in I}\eps_i;
\eeq
    \nhx
     the quantity $\eps_t(I)=1$ if the time ($t$) axis belongs to the
     factor space $\M_I$ and $\eps_t(I) = -1$ otherwise. If $\eps_t(I) =1$,
     this is a true electric or magnetic field, otherwise the
     $F$-form behaves as an effective scalar or pseudoscalar in the
     external subspace (such $F$-forms will be called {\sl quasiscalar\/}).
     The solutions will be written for arbitrary $\eta_s$.


\section{Solutions}  

     Let us use, as in \cite{Br73}, the harmonic
     $u$ coordinate:  $\alpha (u)= d_0\beta^0 + \sigma_1$, \
     $\sigma_1 \eqdef \sumn d_i \beta^i(u)$.

     The Maxwell-like equations for the $F$-forms are easily integrated
     giving
\bear                                                  \label{3.2}
	F_{\e I}^{uM_1\ldots M_{d(I)}}
		\eql Q_{\e I}\e^{-2\alpha - 2\olam_{\e I}\ophi}
                     \ \eps^{M_1...M_{d(I)}}/\sqrt{|g_I|},
		        \qquad  Q_{\e I}= \const,      \\
	F_{\m I,\, M_1\ldots M_{d(\oI)}}                  \label{3.3}
           \eql Q_{\m I}\, \eps_{M_1...M_{d(\oI)}} \sqrt{|g_\oI|},\qquad
		        \qquad  Q_{\m I}= \const,
\ear
     where $|g_I| = \prod_{i\in I} |g^i|$, $Q_s$ are charges
     and bars over $\lambda$ and $\varphi$ denote summing in $a$.
     In what follows we restrict the set $\S=\{s\}$ to such $s$ that
     $Q_s\ne 0$.

     Let us assume (as usual) that
     {\sl neither of our $p$-branes involves the ``external" subspace\/}
     $\M_0$, (that is, $0 \not\in I$ if $\Qei\ne 0$ or $\Qmi \ne 0$).
     Then, if $z$ belongs to $\M_0$, for the total energy-momentum tensor
     (EMT) one has $T_u^u + T_z^z =0 $. The corresponding
     combination of the Einstein equations $R_M^N-\half\delta_M^N R =T_M^N$
     has the form of the Liouville equation giving
\beq
	\e^{\beta^0 - \alpha} =(d_0-1) S(wK_0,\ k,\ u),
	\cm   k = \const,                               \label{3.8}
\eeq
     where one more integration constant (IC)
     is suppressed by choosing the origin of $u$ and
\beq
     \nq S(1,\ h,\ t) = \vars{ h^{-1} \sinh ht, \quad & h>0,\\
     			                        t,       & h=0,\\
     	             	       h^{-1} \sin ht,     & h<0; } \qquad
\begin{array}{rcl}
     S(-1,\ h,\ t)\eql h^{-1} \cosh ht; \quad\  h> 0;            \\
     S(0,\ h,\ t) \eql \e^{ht}, \cm\qquad    h\in \R.
\end{array}                                               \label{3.9}
\eeq

     With (\ref{3.8}) the $D$-dimensional line element may be written in the
     form ($\od \eqdef d_0-1$)
\beq
     ds^2 = \frac{\e^{-2\sigma_1/\od}}{[\od S(wK_0,k,u)]^{2/\od}}
     	  \biggl[ \frac{w\, du^2}{[\od S(wK_0,k,u)]^2} + ds_0^2\biggr]
        		+ \sumn \e^{2\beta^i}ds_i^2.         \label{3.10}
\eeq

    For the remaining field equations let us use
    the so-called $\sigma$-model (minisuperspace) approach (for its more
    general form see \cite{IM5}). Namely, let us treat the set
    of unknowns $\beta^i(u)$, $\varphi^a (u)$ ($i=1,\ldots,n$) as a
    real-valued vector function $x^A (u)$ in an $(n+|\A|)$-dimensional
    vector space $\V$.  Our field equations for $\beta^i$ and $\varphi^a$
    can be derived from the Toda-like Lagrangian
\beq                                                      \label{3.12}
\nqq L=G _{AB}\dot x^A\dot x^B-V_Q (y)
     \equiv \sumn (\dot\beta^i)^2 + \frac{\dot\sigma_1^2}{\od}
	          + \delta_{ab}\dot\varphi^a \dot\varphi^b - V_Q (y),
	  \cm\    V_Q (y) = -\sums \ep_s Q_s^2 \e^{2y_s}
\eeq
	with the ``energy" constraint
\beq                                                      \label{3.16}
	E = G_{AB}\dot x^A \dot x^B + V_Q (y)
	                                =\frac{\od+1}{\od}K, \cm\
        K = \vars {
                      k^2 \sign k, & wK_0 = 1; \\
                      k^2,         & wK_0 = 0,\ -1.       }
\eeq
     where the IC $k$ has appeared in (\ref{3.8}).
     The nondegenerate symmetric matrix
\beq                                                       \label{3.13}
       (G_{AB})=\pmatrix {
  	       d_id_j/\od + d_i \delta_{ij} &       0      \cr
	          0                         &  \delta_{ab} \cr }
\eeq
     defines a positive-definite metric in $\V$;
     the functions $y_s(u)$ are defined as scalar products:
\beq                                                       \label{3.15}
     y_s = \sum_{i\in I}d_i \beta^i - \chi_s \olam_s\ophi
	   \equiv Y_{s,A}  x^A,    \cm\
     (Y_{s,A}) = \Bigl(d_i\delta_{iI_s}, \ \  -\chi_s \lambda_{sa}\Bigr),
\eeq
     where $\delta_{iI} =1$ if $i\in I$ and $\delta_{iI}=0$
     otherwise); the sign factors $\ep_s$ and $\chi_s$ are
\beq
	\ep_{\e I} = -\eta_{\e I} \eps(I), \cm\
	\ep_{\m I} = w \eta_{\m I} \eps(\oI);\cm\      \label{3.14}
	\chi_{\e I}= +1, \cm\
	\chi_{\m I}= -1.
\eeq
     The contravariant components
     and scalar products of the vectors $\vY_s$
     are found using the matrix $G^{AB}$ inverse to $G_{AB}$:
\bearr                                                      \label{3.18}
     (G^{AB}) = \pmatrix{
	\delta^{ij}/d_i - 1/\oD &      0      \cr
	0                       &\delta^{ab}  \cr }, \inch
	(Y_s{}^A) =
  \Bigl(\delta_{iI}-\frac{d(I)}{\oD}, \quad -\chi_s \lambda_{sa}\Bigr); \\
\lal  Y_{s,A}Y_{s'}{}^A \equiv \vY_s \vY_{s'}
	                  = d(I_s \cap I_{s'})                \label{3.20}
     			      - \frac{d(I_s)d(I_{s'})}{\oD}
			      + \chi_s\chi_{s'} \olam_s \olam_{s'}, \cm
     \cm    \oD = D-2.
\ear

\subsection{Orthogonal systems (OS)}  

     The field equations are entirely integrated if all
     $\vY_s$ are mutually orthogonal in $\V$, that is,
\beq                                                          \label{3.21}
     \vY_s \vY_{s'} = \delta_{ss'}\big/ \Nsq, \cm
	     1\big/ \Nsq =
	d(I)\bigl[1- d(I)/\oD \bigr] + {\olam_s}^2 >0.
\eeq
     Then the functions $y_s(u)$ obey the decoupled Liouville equations
     $\ddot y_s = b_s\e^{2y_s}$, whence
\beq                                                   \label{3.23}
     \e^{-y_s(u)} = \sqrt{|b_s|} S(\ep_s,\ h_s,\ u+u_s),
\eeq
     where $b_s \eqdef \ep_s Q_s^2/\Nsq$,
     $h_s$ and $u_s$ are ICs and the function $S(.,.,.)$ has been
     defined in (\ref{3.9}). For the sought functions
     $x^A (u) = (\beta^i,\ \varphi^a)$
     and the ``conserved energy" (\ref{3.16}) we then obtain:
\bear                                                      \label{3.24}
     x^A(u) \eql \sums \Nsq Y_s{}^A y_s(u) + c^A u + \uc^A, \\
          E \eql \sums \Nsq h_s^2\sign h_s + \vec c\,{}^2    \label{3.29}
                     = \frac{d_0}{d_0-1} K.
\ear
     where the vectors of ICs $\vec c$ and $\vec\uc$ are orthogonal
     to all $Y_s$: \ $c^A Y_{s,A} = \uc^A Y_{s,A} = 0$, or
\beq
     c^i d_i\delta_{iI_s} - c^a\chi_s\lambda_{sa}=0, \inch
     \uc^id_i\delta_{iI_s} -\uc^a\chi_s\lambda_{sa}=0.     \label{3.25}
\eeq

\subsection{Block-orthogonal systems (BOS)} 

     One can relax, at least partly, the orthogonality requirement
     (\ref{3.21}), assuming that some of the functions
     $y_s$ (\ref{3.15}) coincide.
     Suppose that the set $\S$ splits into several non-intersecting
     subsets, $\S = \bigcup_{\o}\Som$, $|\Som|=m(\o)$, such that the
     vectors $\vY_{\mu(\o)}$ ($\mu(\o) \in
     S_{\o}$) form mutually orthogonal subspaces $\V_\o$ in $\V$:
\beq
     \vY_{\mu(\o)} \vY_{\nu(\o')} = 0, \cm \o \ne \o'.
		     		        \label{*5}
\eeq
     Suppose, further, that for each $\o$ the functions $y_\mu$,
    $
    	\mu \in \{1,\ldots, m\}= \S_{\o}
    $,
     coincide up to additive constants, which may be then absorbed by
     re-defining the charges $Q_\mu$, so that the expression
    $
    	y_\o(u) \eqdef Y_{\mu(\o), A}x^A
    $
     does not depend on $\mu$.
     This coincidence condition overdetermines the set of
     equations.  However, the consistency relations may be written in terms
     of ICs. Indeed, let us fix $\o$ and suppose that $\nu,\nu',m$
     correspond to this $\o$. Comparing the expressions
    $
    	\ddot y_\o = Y_{\nu,A} \ddot x^A
    $
     with different $\nu$, found from the Lagrange
     equations, we obtain a set of linear algebraic equations for the
     charge factors $q_\nu = \ep_\nu Q_\nu^2$:
\beq
     (\vY_{\nu}-\vY_{\nu'}) \vZ_\o =0,\cm
               \vZ_\o \eqdef \summo q_\mu \vY_\mu,  \label{*3}
\eeq
   \nhx
     which must hold for each pair $(\nu, \nu')$. If the $m$ vectors
     $\vY_\nu$ are linearly independent, all their ends and hence all their
     differences lie in a certain $(m-1)$-dimensional plane in $\V_{\o}$.
     Then there is a nonzero vector $\vZ_{\o}$ whose direction is
     orthogonal to this plane and which therefore satisfies (\ref{*3}).
     If, on the contrary, among $\vY_{\nu}$ there are only $l<m$ linearly
     independent vectors, i.e.,
    $
    	\dim \V_{\o} = l < m,
    $,
     there is in general no vector $\vZ\in \V_{\o}$ orthogonal to all
     their differences, so that \eqs (\ref{*3}) have no nonzero solution,
     unless all the above differences lie in a plane of dimension smaller
     than $l$.

     Therefore the set of \eqs (\ref{*3}) always has a nontrivial solution
     with one free parameter (charge) if all $\vY_{\mu (\o)}$ are linearly
     independent and has, in general, no solution otherwise. (One only
     has to take care of all $q_\mu$ being nonzero; if (\ref{*3}) gives
     some $q_\mu=0$, the consideration may be repeated anew without this
     $\mu$, i.e., with the number of branes reduced by one.)

     Assuming that \eqs(\ref{*3}) have been solved for each $\o$, the
     solution process is completed as in \ssect {2.1}. Define
\bearr
 \nq  \hq_\o = \summo q_\mu; \cm
         b_\o = \vY_{\nu(\o)}\summo q_\mu \vY_\mu;
                                                       \cm
     \vY_\o = \frac{1}{\hq_\o}\summo q_\mu \vY_\mu; \cm
     N_\o^{-2} =
		\vY_\o^2 = \frac{b_\o}{\hq_\o},      \label{*6}
\ear
     where $b_\o$ is $\nu$-independent due to (\ref{*3}).
     (We assume $b_\o \ne 0$, otherwise
     $\ddot y_\o=0$, making this degree of freedom trivial.)
     Then in the case $m=1$ we recover a single
     member of the OS of \ssect {2.1}, with the charge factor $b_\o=
     b_s$, the vector $\vY_\o =\vY_s$ (orthogonal to all others) and
     its norm $N_s^{-2}$. Thus single branes and BOS subsystems
     are represented in a unified way. Replacing $s\ \mapsto \ \o$ in
     (\ref{3.23})--(\ref{3.29}),
     but leaving (\ref{3.25}) unchanged,
     one obtains {\sl a generalized solution for the model
     (\ref{2.1}), valid for any BOS\/}.
     The OS solution of \ssect {2.1} is its special case ($m(\o)=1,
     \forall \o$), therefore in what follows we mostly deal with BOS
     solutions.  (Note that the IC vectors $\vec c$ and $\vec {\uc}$ are,
     even in a BOS, orthogonal to each individual $\vY_s$.)

     The metric has the form (\ref{3.10}), with the function $\sigma_1$
\beq                                                       \label{3.28}
     \sigma_1 = - \frac{d_0-1}{D-2}
     		\sumo N^2_{\o}y_\o(u)
     \summo \frac{q_\mu}{\hq_\o}d(I_\mu) + u\sumn c^i + \sumn \uc^i.
\eeq
     For OS ($\o \mapsto s$) the sum in $\mu$ reduces to $d(I_s)$.

     The OS solution is general for a given set of $d_i$ and $\vY_s$
     and contains $|\S|$ independent charges. The BOS
     solution is special: its number of charges coincides with
     $|\{\o\}|$, the number of subsystems; however,
     we thus gain exact solutions for more general sets
     of input parameters, e.g. a one-charge solution can be obtained for
     actually an arbitrary configuration of branes with linearly
     independent vectors $\vY_\mu$.

     Other integrable cases of the model under study can be found using the
     known methods of solving nontrivial Toda systems, see e.g.
     \cite{IMnew,AA}.

\section{Black holes} 

     The positive energy requirement (\ref{2.25}), fixing the input signs
     $\eta_s$ for Lorentzian models, in the notation (\ref{3.14}) reads
     $\ep_s = \eps_t(I)$; this essentially restricts the
     possible solution behaviour.

     In static, spherically symmetric models%
\footnote
{Cosmological solutions for OS are discussed e.g. in \cite{Greb,IMnew}.}
     $u$ is a radial coordinate ($w=+1$, $\M_0=S^{d_0}$, $K_0 = +1$)
     and time is defined as a factor space among $\M_i$, say, $\M_1$, so
     that  $d_1=1$, $\eps_1 = -1$. The range of $u$ is $(0,\umx)$, where
     $u=0$ corresponds to spatial infinity, while $\umx$ may be finite or
     infinite.  The factor $wK_0$ in (\ref{3.8}) is $+1$, while
     $\ep_s$ is, by (\ref{2.25}), $+1$ for true electric
     and magnetic forms $F_s$ and $-1$ for quasiscalar ones. The general
     solution, combining hyperbolic, trigonometric and power functions for
     various signs of $k$ and $h_\o$, shows a diversity
     of behaviours, but a generic solution has a naked singularity.
     Possible exceptions are \bhs\ (BHs) and \whs.

     BH solutions are obtained in the case $h_\o >0$,
     $\umx=\infty$. The functions
     $\beta^i$ ($i=0,2,\ldots,n$) and $\varphi^a$ remain finite as
     $u\to\infty$ under the following constraints on the ICs:
\bear
     h_\o = k, \cm \forall\  \o; \inch
     c^A = k \sumo \Noq Y_\o{}^A - k \delta^A_1            \label{5.4}
\ear
     where $A=1$ corresponds to $i=1$ (time). The
     constraint (\ref{3.29}) then holds automatically.

     The regularity conditions hold only if $\delta_{1I_s}=1,\ \forall s$,
     so that all our $p$-branes evolve with time
     (even all members of BOS subsystems). This condition
     selects true electric and magnetic $F$-forms, eliminating quasiscalar
     ones ({\sl an analogue of no-hair theorems%
\footnote{A no-hair theorem similar to the present one was obtained for OS
     in \cite{bim97}; in \cite{br95} such a theorem
     in $D$-dimensional dilaton gravity was proved even for cases when no
     solutions were found.}
     }).
     The subfamily (\ref{5.4}) exhausts all BH solutions under our
     assumptions, except the extreme case $k=0$.

     Under the asymptotic conditions
     $\varphi^a \to 0$, $\beta^i \to 0$ as $u\to 0$,
     after the transformation
\bearr
     \e^{-2ku} = 1 - \frac{2M}{r^{\od}},                    \label{5.5}
	         \qquad  \od \eqdef d_0-1, \qquad M  \eqdef k/\od
\ear
     the BH metric and the corresponding scalar fields acquire the form
\bearr
\nq\nhq    ds^2 =
     \biggl(\prod_{\o}H_\o^{A_\o}\biggr)\biggl[-dt^2
     \biggl(1-\frac{2M}{r^\od}\biggr)\prod_\o H_\o^{-2\Noq}
     +
     \biggl(\frac{dr^2}{1-2M/r^\od} + r^2 d\Omega^2_{d_0}\biggr)
       + \sum_{i=2}^{n} ds_i^2
                       \prod_{\o} H_\o^{A_\o^i}\biggr]; \nnnv\label{5.7}
\quad A_\o \eqdef \frac{2}{b_\o}
                  \summo \frac{q_\mu d(I_{\mu})}{\oD}
	                             \eqos 2\Nsq\frac{d(I_s)}{\oD}; \cm
     A_\o^i \eqdef -\frac{2}{b_\o}
                    \summo q_\mu\delta_{iI_\mu}
			       \eqos -2\Nsq \delta_{iI_s};   \yy  \lal
     \varphi^a = \sumo \frac{1}{b_\o} \ln H_\o
		    \summo \chi_\mu q_\mu \lambda_{\mu a}
               \eqos \sums \Nsq \chi_s \lambda_{sa} \ln H_s,
\ear
     where $\eqos$ means ``equal for OS, with $\o=s$";
     $d\Omega^2_{d_0}$ is the line element on $S^{d_0}$
     and $H_\o$  are harmonic functions in $\R_+ \times S^{d_0}$:
\beq                                                        \label{5.8}
	H_\o (r) = 1 + {P_\o}/{r^\od}, \cm
		   P_\o \eqdef \sqrt{M^2 + b_\o/\od^2} - M.
\eeq
     The active gravitational mass $M_{\rm g}$ and
     the Hawking temperature, calculated by standard methods, are
\beq                                                         \label{5.9}
%
	G_{\rm N} M_{\rm g} = M + \sumo \Noq Y_\o^1\, P_\o;
\cm
     \TH = \frac{\od}{4\pi k_{\rm B} (2M)^{1/\od} }
	\prod_\o \biggl(\frac{2M}{2M+ P_\o}\biggr)^{\Noq},
\eeq
     where $G_{\rm N}$ is Newton's constant of gravity.
     The extreme case of minimum mass for given charges $Q_s$
     is $M \to 0$ ($k \to 0$). $\TH$ is zero in the limit $M \to 0$ if
     the parameter $\xi \eqdef \sumo\Noq - 1/\od >0$, is finite if
     $\xi=0$ and is infinite if $\xi <0$. In the latter case%
\footnote{As is explicitly shown in \cite{bim97}, an infinite value of
          $\TH$ can occur only at a curvature singularity.}
     the horizon turns into a singularity as $M \to 0$.

     The behaviour of $\TH$ as $M \to 0$ characterizes the BH
     evaporation dynamics. As $\TH$ depends on $\Noq$, which
     in turn depend on the $p$-brane setup, the latter is
     potentially observable via the Hawking effect.

     Some recent unification models involve several time coordinates (see
     \cite{2times,im2t} and references therein). Our solutions with
     arbitrary signatures $\eps_i$ include all such cases.
     If there is another time direction, $t'$, it is natural to assume that
     some ``branes" evolve with $t'$. However, if we try to find a BH
     with two or more times on equal footing, such that for other times
     $t'$ there is also $g_{t't'}=0$ at the horizon, we have to consider
     $d_1>1$. A calculation then shows that the regularity conditions are
     at variance with the constraint (\ref{3.29}).  We conclude that even
     in a space-time with multiple times a BH can only exist with its
     unique preferred, physical time, while other times are not
     distinguished from extra spatial coordinates.

\section{Lorentzian wormholes and a universal restriction}

     Our solution describes a \wh\ (WH) --- a nonsingular configuration
     with an infinite ``radius" $\e^{\beta^0}$ at both ends of the range of
     $u$ --- if $k<0$ and, in addition, the closest positive zero of
     $S(+1,k,u) \sim \sin |k|u$ (i.e.  $u = \pi/|k|$) is smaller than that
     of $\sin [|h_\o|(u+u_\o)]$ for any $h_\o <0$.  In static, \sph models
     we deal then with a traversable Lorentzian WH; Euclidean WHs are
     closely related to instantons.

     By (\ref{3.29}), for $k<0$ at least some $h_\o$ must be negative as
     well. In a WH solution, for all $\e^{y_\o}$ to be regular,
     we must have $|k| >|h_\o|$ for all $h_\o <0$. Furthermore, for $k<0$
     and $h_\o<0$ one needs $wK_0=1$ and $b_\o > 0$, respectively.

     Consider, for simplicity, the OS solution, so that $b_\o =  b_s$. The
     following table shows the sign factors $wK_0$
     and $\ep_s=\sign b_s$ for forms with positive energy in different
     models.

\bigskip            \label{tab}

\begin{tabular}{||l|l|c|c|c||}
\hline
\multicolumn{2}{||c|}{}  & Cosmology&  Static spaces  &  Euclidean \mas \\
\multicolumn{2}{||c|}{}  &   $w=-1$ &    $w=+1$       &    $w=+1$  \mas \\
\hline
\multicolumn{2}{||c|}{$wK_0$}& $-K_0$ &  $K_0$       &    $K_0$ \mas \\
\hline
	    & electric    &  none    &     $+ 1$       &    none  \mas \\
\cline{2-5}
\raisebox{-1ex}[0ex][0ex]{$\ep_s$}
	    & magnetic    &  none    &      $+ 1$      &    none  \mas \\
\cline{2-5}
	    & electric quasiscalar
	                  &  $- 1$   &    $- 1$        &     $- 1$  \mas \\
\cline{2-5}
	    & magnetic quasiscalar
	                  &   $- 1$  &    $- 1$        &     $+ 1$  \mas \\
\hline
\end{tabular}
\bigskip

     We see that WHs can exist in static or Euclidean models with
     spherical symmetry rather than pseudospherical or planar one.  In
     static models one needs true electric and magnetic fields. Both in
     cosmology and in Euclidean models all $F$-forms are quasiscalar since
     time is the $u$ coordinate, out of any $I$, but in cosmology no fields
     are able to create $h_s<0$. The Wick rotation from Lorentzian
     cosmology, preserving all $\eta_s$, changes $w$ and hence
     $\ep_{\m I}$, leaving the same $\ep_{\e I}$.  This distinction is
     related to the property of the duality transformation to change the
     EMT sign in Euclidean models \cite{GiSt}.

     Suppose $k<0$. Since in \eq (\ref{3.29}) $\vec c\,{}^2\geq 0$,
     the requirement $|k| > |h_s|$ means that
\beq                                                   \label{4.2a}
     \sum_{\{s:\ h_s <0\}}  \Nsq > \frac{d_0}{d_0-1}
\eeq

     This inequality is not only {\sl necessary\/}, but also
     {\sl sufficient\/} for the existence of WHs with given input
     parameters: $d_i$ and the vectors $\vY_s$. Indeed,
     put $c^A=0$ and $Q_s=0$ for all quasiscalars and choose all
     $h_s<0$ to be equal, then due to (\ref{4.2a}) $|h_s| < |k|$.  It is
     now easy to choose the ICs $u_s$ so that $\sin [|h_s|(u+u_s)] > 0$ on
     the whole segment $[0,\pi/|k|]$ --- and this yields a WH.

     On the other hand, in $(d_0+2)$-dimensional general relativity,
     in the neighbourhood of a WH throat, matter must violate the
     weak energy condition (see e.g. \cite{HV}).
     In a $(d_0+2)$-dimensional formulation of the present model, with
     or without $\varphi^a$, one can prove that under the
     condition (\ref{2.25}) the
     weak energy condition holds (see more details in \cite{br-new}),
     so that Lorentzian WHs can appear as well only
     at the expence of explicitly invoking negative energies%
\footnote{Indeed, if at least one of $\varphi^a$ is pure imaginary,
          its $c^a$ is also pure imaginary, and, as is clear
          from (\ref{3.29}), wormhole solutions are readily obtained
          (cf. \cite{Br73,b}). Cancelling the positive energy requirement
	  for quasiscalar $F$-forms leads to the same result.}.

     The non-existence of spherical WHs is incompatible with
     (\ref{4.2a}), and the properly formulated opposite inequality must hold:
\beq
   \nq \sums \delta_{1I_s}\Nsq \leq \frac{d_0}{d_0-1},    \label{4.7a}
	\cm \mbox{or for $\lambda_{sa}=0$:}\qquad
     \sums \delta_{1I_s} \biggl[ d(I_s)\biggl(1
	        \frac{d(I_s)}{D-2}\biggr)\biggr]^{-1}
	                                  \leq \frac{d_0}{d_0-1},
\eeq
     where the factor $\delta_{1I_s}$ excludes quasiscalars. A more
     general formulation of this conclusion is

\medskip\noi
{\bf Statement 1.}\
     {\sl Given a vector space $\V$ with the metric
     (\ref{3.13}), where $d_i$, $i=0,\ldots,n$, are positive integers,
     $d_0>1$, $d_1=1$, $\oD=\sum_{i=0}^{n}d_i-1$. Then for any set of
     vectors $\vY_s$ defined in (\ref{3.15})
     ($I_s \subseteq \{1,\ldots,n\}$, $\chi_s\lambda_{sa}\in\R$),
     mutually orthogonal in $\V$, the inequality (\ref{4.7a}) is valid.\/}
\medskip\noi

     This formulation does not mention $F$-forms, time, etc., and is
     actually of purely combinatorial nature.
     No doubt there exists its combinatorial proof,
     but, surprisingly, it has been obtained here from physically
     motivated analytical considerations.

     Statement 1 is valid for any set of $F_s$-forms under the
     specified conditions, even if this set is only a subsystem in a
     bigger model, for which maybe we do not know any solution.

     All this can be repeated for BOS with certain complications. In
     (\ref{4.2a}) one should then substitute $s \mapsto \o$,
     while the first inequality (\ref{4.7a}) will read:
    $
     \sumo \delta_{1I_{\mu(\o)}}\Noq \leq d_0/(d_0-1)
    $.

\section{Euclidean wormholes} 

     As seen from the table in p.{\pageref{tab}},
     in Euclidean models with the action
     (\ref{2.1}) WH solutions can be built only with the aid of
     magnetic forms $F_s$. All $F$-forms are now quasiscalar and we are
     no more restricted to $I_s$ containing a distinguished index. The
     condition (\ref{4.2a}) (or its BOS version) is, as before, necessary
     and sufficient for WH existence, but there is a wider choice of $I_s$
     able to give $h_s<0$ or $h_\o<0$ and, as a result, to fulfil it.

     Classical Euclidean WHs are used to describe
     quantum tunneling, where the finite action plays a crucial
     role. Let us find it for WHs appearing among our solutions.

     The Euclidean action $\SE$ corresponding to (\ref{2.1}) is written as
\beq                                                   \label{E1}
     \SE = \frac{1}{2\kappa^{2}}
                  \int\limits_{\M} d^{D}z \sqrt{g} \biggl\{
     -R[g]
     + \delta_{ab} g^{MN} \d_{M} \varphi^a \d_{N} \varphi^b
                                                 + \sum_{s\in \S}
     \frac{1}{n_s!} \e^{2 \lambda_{sa} \varphi^a} F_s^2
               \biggr\},
\eeq
     where, for simplicity, we put $\eta_s=1$ and all
     $\eps_i=+1$ (full Euclidean signature).
     Due to the trace of the Einstein equations, $R[g]= -T^M_M/(D-2)$,
     the curvature cancels the scalar term in the action and the
     remainder is expressed in terms of $F_s^2$.

     WH solutions can contain both electric and magnetic forms and
     for both one has
     $(1/n_s!) \e^{2\alpha+2\olam\ophi}
     	        \, F_s^2  = Q_s^2 \e^{2y_s}$ (within each BOS
	subsystem all $y_s=y_\o$). As a result,
\beq
     \SE = \frac{1}{2\kappa^{2}}                           \label{E2}
                  \int\limits_{\M} d^{D}z
		         \biggl(\prod_{i=1}^{n}\sqrt{g^i}\biggr)
	 \sums \frac{2n_s-2}{D-2}\, Q_s^2 \e^{2y_s}
\eeq


     Let us assume that
(i)  there are only BOS subsystems with $h_\o <0$
     (or, in an OS, only magnetic forms $F_s$ with $h_s <0$);
(ii) the WH is symmetric [i.e. $y_\o(v)=y_\o(-v)$ for all $\o$,
	$v \eqdef u-\pi/(2|k|)$] and
(iii) the boundary condition $y(a){=}y(-a)=0$ is valid,
     where $2a{=}\pi/|k|$ (this is just normalization of charges,
     with no effect on generality). Then the solution (\ref{3.23}) for
     $y_\o$ can be written in the form
%
     $\e^{-y_\o} = (\sqrt{b_\o}/|h_\o|) \cos |h_\o|v$, where $b_\o >0$.

     The extra dimensions, provided their volumes are finite, can be
     integrated out in (\ref{E2}). The remaining integrals in $v$ are
     easily calculable. The final expression is
\beq                                                         \label{E7}
     \SE =
     \frac{1}{16\pi G_{\rm N}}\frac{2\pi^{\od/2 +1}}{\Gamma(\od/2 +1)}
	\sumo
		\frac{2}{b_\o}\sqrt{b_\o - h_\o^2} \biggl(
	\summo
		\frac{2n_\mu -2}{D-2}Q_\mu^2 \biggr).
\eeq
     where, for correspondence, the factor $1/2\kappa^2$
     times the internal space volumes ($i= 1,\ldots,n$) is identified
     with $(16\pi G_{\rm N})^{-1}$ and the second factor in (\ref{E7})
     is the volume of $S^{d_0}$ ($\Gamma$ is Euler's gamma function).

\def\rth{r_{\rm th}}

     Another quantity of interest is the radius $\rth$ of the WH throat,
     determining a length scale. For a symmetric WH it is
     just the value of $\e^{\beta_0}$ at $v=0$. Under the above
     assumptions,
\beq                                                        \label{E8}
	\rth = \biggl(\frac{|k|}{\od}\biggr)^{1/\od}
	\prod_\o [y_\o (0)]^{A_\o/2},
\cm
	y_\o(0) = \sqrt{b_\o}/|h_\o|,
\eeq
     where $A_\o$ has been defined in (\ref{5.7}).
     Simplifications for OS are evident.

     Quantum versions of the present models can be obtained as described
     e.g. in \cite{IMnew}.

\section{Examples}

{\bf 1.} The simplest and even degenerate example
     of $F_s$ is the Kalb-Ramond field
     $F_{MNP}$ in 4 dimensions, where the indices do not contain $u$.
     It is a magnetic form with $I=\emptyset$, the coupling
     $\lambda=0$, so that its vector $\vY=0$, and the solution containing
     only $F_{MNP}$ is trivial.  The Euclidean WH solution obtained in this
     case, with its $\SE$ and $\rth$, coincides with that of
     Refs.\,\cite{GiSt} after adjusting the notations.

\medskip\noi
{\bf 2.} In 11-dimensional supergravity \cite{brane}, with
     $\varphi^a = \olam_s =0$, the orthogonality conditions (\ref{3.21})
     are satisfied by 2-branes, $d(I_s)=3$, and 5-branes, $d(I_s)=6$,
     if the ``intersection rules" hold:
     $d(3\cap 3) =1$, $d(3\cap 6)=2$, $d(6\cap 6)=4$ (the notations are
     evident).  In particular, with $d_0=2$ or $d_0=3$ and other $d_i=1$,
     there can be seven mutually orthogonal 2-branes, but no more than
     three of them can have $\delta_{1I_s}=1$, i.e., describe true electric
     or magnetic forms in a static space-time \cite{Greb,bim97}. Since for
     all 2- and 5-branes $\Nsq=1/2$, the BH temperature (\ref{5.9}) in the
     extreme limit tends to infinity if there is one such brane, to a
     finite limit if there are two and to zero if there are three branes.
     Lorentzian WHs are absent since (\ref{4.2a}) requires $\sums\Nsq >2$
     for $d_0=2$ and $>3/2$ for $d_0=3$. In the Euclidean case we can have
     as many as 7 magnetic 2-branes, each with $\Nsq=1/2$, and WHs are
     easily found.

\medskip\noi
{\bf 3.} Example of a BOS in the same model: let the digits $1,\ldots,7$
     label 1-dimensional internal spaces, and consider a set of 6 branes
     with the following $I_s$:\yy  \hspace*{1cm}
$
	\begin{array}{ll} \dst
	a: & 123456, \\
	b: & 123,
	\end{array} \cm
		 	\begin{array}{ll} \dst
				c: & 234,\\
        			d: & 147,
	        	\end{array} \cm
			    	           \begin{array}{ll}\dst
					   e: & 257,\\
					   f: & 367.   \end{array}\yy
$
    The $F$-forms associated with $a,b,c$, with ``wrong" intersections
    (3- and 2-dimensional), form a BOS subsystem; $d,e,f$ are separate
    forms with $\vY_s$ orthogonal to all others. In this example our scheme
    gives (as it seems) a new solution with 4 charges.

\Acknow
 {I am grateful to V. Ivashchuk, V. Melnikov and J. Fabris for many helpful
 discussions, to CAPES (Brazil) for partial financial support, and to
 colleagues from DFis-UFES, Vit\'oria, for kind hospitality.}

\end{document}